\def \VersionAuthor {}
\documentclass[EPiC]{easychair}

\makeatletter
\AtBeginDocument{%
  \@ifpackageloaded{hyperref}
  {\def\@doi#1{\href{https://doi.org/#1}
      {\ttfamily https://doi.org/#1}\egroup}}
  {\def\@doi#1{\ttfamily https://doi.org/#1\egroup}}
  \def\doi{\bgroup\catcode`\_=12\relax\@doi}}
\makeatother
\usepackage[ruled,vlined,linesnumbered]{algorithm2e}
	\SetKwInOut{Input}{input}
	\SetKwInOut{Output}{output}
\usepackage[utf8]{inputenc}
\usepackage[english]{babel}
\usepackage{booktabs}

 \usepackage{graphicx}
 \usepackage{amssymb,amsmath}%
 
\usepackage[misc,geometry]{ifsym} %

\ifdefined\VersionAuthor
	\usepackage[backend=biber,backref=true,style=alphabetic,url=false,doi=true,defernumbers=true,sorting=anyt,maxnames=99]{biblatex} %
	\renewbibmacro*{doi+eprint+url}{%
		\iftoggle{bbx:doi}
			{\color{black!40}\footnotesize\printfield{doi}}
			{}%
		\newunit\newblock
		\iftoggle{bbx:eprint}
			{\usebibmacro{eprint}}
			{}%
		\newunit\newblock
		\iftoggle{bbx:url}
			{\usebibmacro{url+urldate}}
			{}%
	}

\fi
\ifdefined\VersionWithComments
	\usepackage{draftwatermark}
	\SetWatermarkText{draft}
	\SetWatermarkScale{15}
	\SetWatermarkColor[gray]{0.9}
\fi
\usepackage[capitalise,english,nameinlink]{cleveref} %
\crefname{line}{\text{line}}{\text{lines}} %

\usepackage{tikz}
\usetikzlibrary{decorations.pathmorphing} 
\usetikzlibrary{arrows,calc}
\tikzset{
>=stealth',
help lines/.style={dashed, thick},
axis/.style={<->},
important line/.style={thick},
connection/.style={thick, dotted},
}

\ifdefined\VersionWithComments
	\usepackage[colorinlistoftodos,textsize=footnotesize]{todonotes}
\else
	\usepackage[disable]{todonotes}
\fi
\newcommand{\gennote}[3]{\todo[linecolor=#2,backgroundcolor=#2!25,bordercolor=#2]{#3: #1}\xspace}
\newcommand{\ea}[1]{\gennote{#1}{blue}{ÉA}}

\newcommand{\jj}[1]{{\gennote{#1}{purple}{Jawher}}}
\ifdefined \VersionWithComments
	\newcommand{\todoinline}[1]{\mbox{}{\color{red}{\textbf{TODO}\ifx#1\\\else:\ \fi #1}}} %
\else
	\newcommand{\todoinline}[1]{}
\fi

\usepackage{verbatim} %

\ifdefined \VersionWithComments
	\usepackage{soul}
	\newcommand{\reviewAdd}[1]{{\color{purple}#1}}
	\newcommand{\reviewDelete}[1]{{\color{red}\st{#1}}}
\else
	\newcommand{\reviewAdd}[1]{#1}
	\newcommand{\reviewDelete}[1]{}
\fi

\ifdefined\VersionAuthor
\else
	\usepackage[pagewise]{lineno} %
	\linenumbers
	
\fi

\ifdefined \VersionWithComments
 	\definecolor{colorok}{RGB}{80,80,150}
\else
	\definecolor{colorok}{RGB}{0,0,0}
\fi

\newcommand{\eg}{\textcolor{colorok}{e.\,g.,}\xspace}

\newcommand{\ie}{\textcolor{colorok}{i.\,e.,}\xspace}

\RequirePackage{amsmath}

\usepackage{mathabx}

\usepackage{color}

\makeatletter
\newcommand\footnoteref[1]{\protected@xdef\@thefnmark{\ref{#1}}\@footnotemark}
\makeatother

\usepackage{amsthm}
\theoremstyle{plain}

\newtheorem{proposition}{Proposition}

\theoremstyle{definition}
\theoremstyle{remark}
\newtheorem{remark}{Remark}

\title{Guaranteed phase synchronization of  hybrid oscillators using symbolic Euler's method: %
The Brusselator and biped examples
}
\author{%
Jawher Jerray \inst{1}
\and
Laurent Fribourg \inst{2}
\and
Étienne André \inst{3}
}

\institute{$^1$	Université Sorbonne Paris Nord, LIPN, CNRS, UMR 7030, F-93430, Villetaneuse, France \\
	$^2$ Université Paris Saclay, LSV, CNRS, ENS Paris Saclay\\
	$^3$ Université de Lorraine, CNRS, Inria, LORIA, F-54000 Nancy, France\\
  \email{jerray@lsv.ens-cachan.fr }
}

\graphicspath{{./figures/}}

\titlerunning{Phase synchronization using Euler's method}

\authorrunning{L. Fribourg, J. Jerray} 

\begin{document}
\sloppy

\maketitle

\begin{abstract}

The phenomenon of {\em phase synchronization} was evidenced in the 17th century by 
Huygens while observing two pendulums of clocks leaning against the same wall. This phenomenon has more recently appeared as a widespread phenomenon in nature, and turns out to have multiple industrial applications.
The exact parameter values of the system
for which the phenomenon manifests itself are however delicate to obtain in general, and it is interesting to find formal sufficient conditions to {\em guarantee} phase synchronization.
Using the notion of {\em reachability}, we give here such a formal method. More precisely, our method selects a portion $S$ of the state 
space, and shows that any solution starting at $S$ {\em returns} to~$S$ within
a fixed number of periods \reviewDelete{$T$}\reviewAdd{$k$}. Besides, our
method shows that the components of the solution are then (almost) in phase.
We explain how the method applies on the
Brusselator reaction-diffusion and the biped walker examples.

\end{abstract}

\section{Introduction}

The phenomenon of phase synchronization was evidenced in the 17th century by 
Huygens while observing two pendulums of clocks leaning against the same wall. This phenomenon has more recently appeared as a widespread phenomenon in nature, and turns out to have multiple industrial applications \cite{Winfree:1980,Mirollo,Kiss1676,Acebron05}.

Basically, we consider a system consisting of two periodic coupled oscillators. After a certain time, the same period $T$ for both oscillators is found, and, whatever the initial condition of each oscillator, the two components evolve in phase on their respective orbits.

The exact parameter values of the system
for which the phenomenon manifests itself are however delicate to obtain in general, and it is interesting to find formal sufficient conditions to guarantee phase synchronization.
There is a classical method, called ``direct'', which is used to characterize such conditions \cite{Winfree:1980}.
Basically, this method starts from a pair 
of synchronized components evolving on their respective orbits, then moves ``slightly'' apart each component   (with the help of a small perturbation), and observes, after a fixed number of periods, say $k$, that the phases of the two components have become very close to each other again (see \eg{}  \cite[Appendix H]{Nakao17} for a formal description). Such a  method shows besides that the synchronization is {\em robust} (or ``stable'') since, after a small disturbance, the system resynchronizes quickly (see, \eg{} \cite{Maginu79}).

We will reproduce the spirit of this method using the notion of {\em reachability}. More precisely, our method selects a portion $S$ of the state 
space, and shows that any solution starting at~$S$ {\em returns} to~$S$ within
a fixed number of periods~\reviewDelete{$T$}\reviewAdd{$k$}. Besides, our
method shows that the components of the solution are then (almost) in phase.

After a formal description of the method, we explain how the method applies
on the
Brusselator reaction-diffusion and the biped walker examples.

\paragraph{Plan}
In \cref{section:method}, we explain the underling principle of our method, which is based on the
notion of reachability. We describe in \cref{section:Euler} how this principle is implemented using symbolic Euler's method. We illustrate the method
on the Brusselator reaction-diffusion example (\cref{section:brusselator}) and the biped walker
example (\cref{section:biped}).
We conclude in \cref{section:conclusion}.

\section{Showing synchronization using a reachability method}\label{section:method}

We consider a system composed of $n$ subsystems governed
by a system of differential equations (ODEs)
of the form
$\dot{x}(t)=f(x(t))$. For the sake of simplicity, we suppose here $n=2$.\footnote{The extension of the method to $n\geq 3$ is straightforward in principle, but is a source of combinatorial explosion.}
The system of ODEs is thus  of the form:
\begin{align*}
\dot{x_1}(t) &= f_1(x_1(t),x_2(t))
\\
\dot{x_2}(t) &= f_2(x_1(t),x_2(t))	
\end{align*}
with $x(t)=(x_1(t),x_2(t))\in \mathbb{R}^m\times \mathbb{R}^m$,
where $m$ is the dimension of the state space of each subsystem.
The initial condition is of the form
$(x_1^0,x_2^0)\in\mathbb{R}^m\times \mathbb{R}^m$.

The set $S=S_1\times S_2$ (with $S_i\subset \mathbb{R}^m$, $i=1,2$) on which we focus our analysis, is selected roughly speaking as follows. We first consider, for each subsystem $i$ ($i=1,2$), a ``ring'' of reduced width $e_i$ around the cyclic trajectory (orbit). We then select a fragment of each ring, which gives two sets of states $S_1$ and $S_2$. Typically, for $i=1,2$, $S_i$ is a  {\em parallelogram}\ea{répétition par rapport à la précédente footnote ?}\jj{footnote 2 à supprimer}
with a {\em horizontal} ``base'' of width $e_i$ (or symmetrically a vertical side).  The set  $S_i$ is thus characterized by a triple $(a_i,b_i,e_i)$ where $a_i$ and $b_i$ are the end points of its main diagonal,
and  $e_i$ the size of its horizontal base.\footnote{The precise finding of the coordinates of $a_i$ and $b_i$, and size $e_i$ ($i=1,2$) for which our method of synchronization applies successfully, is actually a basic difficulty of the method, but this issue is beyond the scope of this paper. We assume here that $a_i, b_i$ and $e_i$ are given.}
We assume that the parallelogram $S_i$ is ``long'', i.e.:

\begin{center}
(H) The width $e_i$ of $S_i$ is ``small'' w.r.t.\ 
$f_i=|ord(b_i)-ord(a_i)|$.
\end{center}

Typically, we have: $e_i/f_i < 1/20=0.05$.
We now consider a point $x^0=(x_1^0,x_2^0)\in S$ 
(\ie{} $x_1^0\in S_1$ and $x_2^0\in S_2$), and consider the following procedure
$PROC0(x^0)$:

\begin{enumerate}
	\item 
Show that, if $x(0)=x^0$,
then there exists $t\in [k T, (k+1) T)$: 
$x(t)\in S$ (\ie{} $(x_1(t), x_2(t)) \in S_1\times S_2$) ({\em recurrence} of $S$), and

\item At $t$, the two components $x_1(t)$ and $x_2(t)$ of $x(t)$ are practically in phase, i.e.: $|\phi(x_1(t))-\phi(x_2(t))| < \epsilon$  ({\em synchronization})
\end{enumerate}

\begin{remark}
IN $PROC0$, we assume that $T,k,\epsilon$ are given constants\ea{c'est un peu flou pour moi à ce stade ce que sont $k$ et $T$ ($\epsilon$, c'est plus clair)}\reviewAdd{, where $T$ is the period and $k$ is the number of periods}.
\end{remark}

\begin{remark}
The procedure guarantees  only  a {\em recurrent} form of synchronization at times $t, t',\dots, t^{(n)},\dots$  with  $nkT \leq t^{(n)} <n(k+1)T$. This is weaker than {\em standard} synchronization which states that, after the end of the perturbation, the state $x(t)$ {\em converges} to a solution whose components are in phase.
\end{remark}

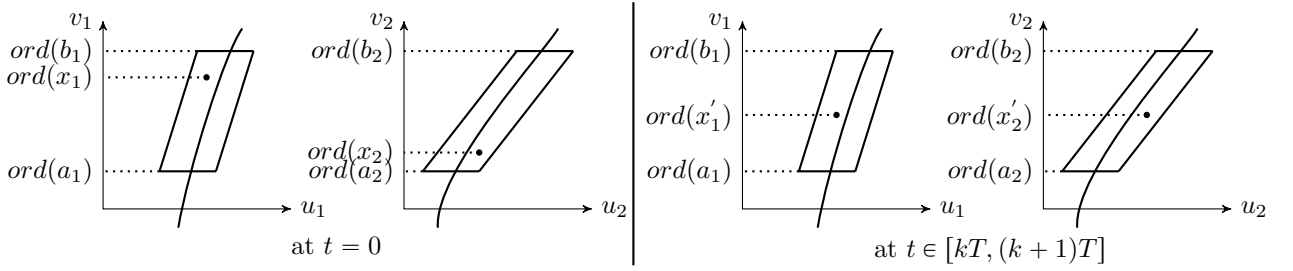
\begin{figure}[t]
  \centering
    \begin{tikzpicture}[scale=0.5]
    \coordinate (y) at (0,5);
    \coordinate (x) at (5,0);
    \draw[<->] (y) node[left] {$v_1$} -- (0,0) --  (x) node[right] {$u_1$};
    \path
    coordinate (start) at (3.7,4.8)
    coordinate (c1) at (3.25,4.2)
    coordinate (c2) at (2.25,1.)
    coordinate (slut) at (2.,-.5)
    coordinate (top) at (2.75,3.5)
    coordinate (rect1) at (2.5,4.2)
    coordinate (rect2) at (4,4.2)
    coordinate (rect3) at (3,1.)
    coordinate (rect4) at (1.5,1.);
    \draw[important line] (start) .. controls (c1) and (c2) .. (slut);
    \draw[important line] (rect1) -- (rect2) (rect2) -- (rect3) (rect3) -- (rect4) (rect4) -- (rect1);
    \filldraw [black] (top) circle (2pt) node[above right, black]{};
    \draw[connection] (rect1)--(0,4.2) node[left] {$ord(b_1)$};
    \draw[connection] (top)--(0,3.5) node[left] {$ord(x_1)$};  
    \draw[connection] (rect4)--(0,1) node[left] {$ord(a_1)$};
    
    \begin{scope}[xshift=8cm]
        \coordinate (y) at (0,5);
        \coordinate (x) at (5,0);
        \draw[<->] (y) node[left] {$v_2$} -- (0,0) --  (x) node[right] {$u_2$};
        \path
        coordinate (start) at (4.1,4.8)
        coordinate (c1) at (3.75,4.2)
        coordinate (c2) at (0.8,1.)
        coordinate (slut) at (0.9,-.5)
        coordinate (top) at (2.,1.5)
        coordinate (rect1) at (3.,4.2)
        coordinate (rect2) at (4.5,4.2)
        coordinate (rect3) at (2,1.)
        coordinate (rect4) at (0.5,1.)
        coordinate (bar_top) at (6.1, 5.5)
        coordinate (bar_bottom) at (6.1, -1.5);
        \draw[important line] (start) .. controls (c1) and (c2) .. (slut);
        \draw[important line] (rect1) -- (rect2) (rect2) -- (rect3) (rect3) -- (rect4) (rect4) -- (rect1);
        \draw[important line] (bar_top) -- (bar_bottom);
        \filldraw [black] (top) circle (2pt) node[above right, black]{};
        \draw[connection] (rect1)--(0,4.2) node[left] {$ord(b_2)$};
        \draw[connection] (top)--(0,1.5) node[left] {$ord(x_2)$};  
        \draw[connection] (rect4)--(0,1) node[left] {$ord(a_2)$};
    \end{scope}
    
    \node [below, text width=10cm,align=justify] at (15,-0.5) {at $t = 0$};
    \begin{scope}[xshift=17cm]
    \coordinate (y) at (0,5);
    \coordinate (x) at (5,0);
    \draw[<->] (y) node[left] {$v_1$} -- (0,0) --  (x) node[right] {$u_1$};
    \path
    coordinate (start) at (3.7,4.8)
    coordinate (c1) at (3.25,4.2)
    coordinate (c2) at (2.25,1.)
    coordinate (slut) at (2.,-.5)
    coordinate (top) at (2.5,2.5)
    coordinate (rect1) at (2.5,4.2)
    coordinate (rect2) at (4,4.2)
    coordinate (rect3) at (3,1.)
    coordinate (rect4) at (1.5,1.);
    \draw[important line] (start) .. controls (c1) and (c2) .. (slut);
    \draw[important line] (rect1) -- (rect2) (rect2) -- (rect3) (rect3) -- (rect4) (rect4) -- (rect1);
    \filldraw [black] (top) circle (2pt) node[above right, black]{};
    \draw[connection] (rect1)--(0,4.2) node[left] {$ord(b_1)$};
    \draw[connection] (top)--(0,2.5) node[left] {$ord(x^{'}_1)$};  
    \draw[connection] (rect4)--(0,1) node[left] {$ord(a_1)$};
    
    \begin{scope}[xshift=8cm]
        \coordinate (y) at (0,5);
        \coordinate (x) at (5,0);
        \draw[<->] (y) node[left] {$v_2$} -- (0,0) --  (x) node[right] {$u_2$};
        \path
        coordinate (start) at (4.1,4.8)
        coordinate (c1) at (3.75,4.2)
        coordinate (c2) at (0.8,1.)
        coordinate (slut) at (0.9,-.5)
        coordinate (top) at (2.75,2.5)
        coordinate (rect1) at (3.,4.2)
        coordinate (rect2) at (4.5,4.2)
        coordinate (rect3) at (2,1.)
        coordinate (rect4) at (0.5,1.);
        \draw[important line] (start) .. controls (c1) and (c2) .. (slut);
        \draw[important line] (rect1) -- (rect2) (rect2) -- (rect3) (rect3) -- (rect4) (rect4) -- (rect1);
        \filldraw [black] (top) circle (2pt) node[above right, black]{};
        \draw[connection] (rect1)--(0,4.2) node[left] {$ord(b_2)$};
        \draw[connection] (top)--(0,2.5) node[left] {$ord(x^{'}_2)$};  
        \draw[connection] (rect4)--(0,1) node[left] {$ord(a_2)$};
    \end{scope}
    \node [below, text width=10cm,align=justify] at (13.5,-0.5) {at $t \in [kT, (k+1)T]$};
    \end{scope}
    
    \end{tikzpicture}
  \caption{Scheme of $S_1$ (left) and $S_2$ (right) at $t=0$ (top)
and for some $t\in [kT, (k+1) T)$ (bottom).}
\label{fig:Pouchol0}
\end{figure}

The notion of {\em phase} $\phi(x_i(s))$, for $i=1,2$  of component $x_i(s)$
at time $s$, remains to be defined in this framework. From a general point of view, one can suppose that, during its traversal of $S_i$, the phase of the point $x_i(s)$ varies, after normalization, between 0 and 1. As $S_i$ is of small dimension with respect to the orbit of the subsystem~$i$, we can assimilate the trajectory described by $x_i(s)$ in $S_i$ to a straight line segment whose ordinate varies 
from $ord(a_i )$ to $ord(b_i)$. Moreover, we can assume that on this small fragment of orbit, the {\em phase velocity} is {\em constant}. 
Given a point of $x_i(s)$ of $S_i\equiv (a_i,b_i,e_i)$ at time $s$ ($i=1,2$),
we can thus define its {\em phase} (in a ``linearized'' and ``normalized'' manner w.r.t.\ $S_i$)  by:
	$$\phi[x_i(s)] = (ord(x_i(s))-ord(a_i))/(ord(b_i)-ord(a_i)),$$
where $ord(x_i(s))$ denotes the ordinate of $x_i(s)$.
See \cref{fig:Pouchol0}.

\section{Symbolic reachability using Euler's method}\label{section:Euler}

The above procedure $PROC0$ takes a {\em point} of $S$ as input. So it is not possible to prove the synchronization of {\em all} the points starting at~$S$, since they are in infinite number. We thus need to consider a {\em symbolic} (or ``set-based'') version of $PROC0$ which takes a {\em dense subset of points} as input. Such subsets are considered here under the form of  ``(double) ball'' 
of the form $B=B_1\times B_2$,
where $B_i\subset \mathbb{R}^m$ ($i=1,2$) is a ball of the form ${\cal B}(c_i,r)$
with $c_i\in \mathbb{R}^m$ ({\em centre}) and $r$ a positive real ({\em radius}).\footnote{$x_i\in {\cal B}(c_i,r)$ means $\|x_i-c_i\|\leq r$ where $\|\cdot\|$ is the Euclidean norm.}

\smallskip

Let $B^0={\cal B}(c_1^0,r^0)\times {\cal B}(c_2^0,r^0)\subset \mathbb{R}^{m}\times  \mathbb{R}^m$, with $c_i^0\in\mathbb{R}^m$ ($i=1,2$)
and $r^0$ positive real.
As a symbolic method, we use here the {\em symbolic Euler's method} \cite{SNR17,F-formats17}
in order to compute (an overapproximation of) the set of solutions starting
at $B^0$.
We define  for $t\geq 0$:
$$B^{euler}(t) = {\cal B}(c_1(t),r(t))\times {\cal B}(c_2(t),r(t)),$$
where $(c_1(t),c_2(t))\in\mathbb{R}^m\times\mathbb{R}^m$ is 
the approximated value of solution $x(t)$ of $\dot{x}=f(x)$
with initial condition $x(0)=(c_1^0,c_2^0)$ given
by {\em Euler's explicit method}, and
$r(t)\approx r^0 e^{\lambda t}$ is the {\em expanded} radius using the
{\em one-sided Lipschitz constant} $\lambda$ (also called ``logarithmic norm'' or ``matrix norm'')
\cite{Soderlind06,AminzareS12}) associated to $f$ (see \cite{F-formats17} for details).\footnote{The value of $\lambda$ is defined ``locally'', and varies according to the position of  $x(t)=(x_1(t),x_2(t))$ in the state space. For regions where $\lambda <0$, the value of $r(t)$ is considered to be constant; the value of $r(t)$ increases only when $x(t)$ occupies a region where $\lambda> 0$ (which corresponds in \cref{fig:Pouchol2lambda} in case $x_1(t)$ or $x_2(t)$ is located in the {\em red} part of its orbit). See \cite{F-formats17}.}
It is shown in \cite{SNR17} that $B^{euler}(t)$ 
contains all the solutions $x(t)$ that start
at $B^0$:
$$B^{euler}(t) \supseteq 
\{x(t)\ | \ x(0)\in B^0\}\equiv  \{(x_1(t),x_2(t))\ |\ (x_1(0),x_2(0))\in {\cal B}(c^0_1,r^0)\times{\cal B}(c^0_2,r^0)\}.\ (*)$$

Given a ball $B=B_1\times B_2\subset\mathbb{R}^m\times\mathbb{R}^m$, the symbolic version  of $PROC0$ is defined as follows:\\

$PROC1(B)$

Let $B^0 := B$. Show that there exists $t \in [kT, (k+1)T)$: 

	1'. 	$B^{euler}(t)\subset S$, i.e.: ${\cal B}(c_i(t),r(t))\subset S_i$ for $i=1,2$.	\ ({\em recurrence}) 

	 2'. 	$|phase(c_{1}(t)) - phase(c_{2}(t))| \leq \epsilon$ 										({\em synchronization})\\

Note that, since ${\cal B}(c_i(t),r(t))\subset S_i$ ($i=1,2$) by (1'), we have:
	$$r(t) \leq \frac{1}{2}\min(e_1,e_2)		\hspace*{4mm} (**)$$
where  $e_i$ denotes the width of $S_i$.\\

\begin{remark}
Works by Aminzare, Sontag, Arcak and others make use of logarithmic norms
to prove phase synchronization 
but only in a {\em contractive} context ($\lambda <0$)  
\cite{Arcak11,AminzareS14,SAAS13}. On the other hand,
logarithmic norms (with possibly $\lambda>0$)
have been used to the symbolic control of hybrid systems \cite{ReissigR19,RunggerR17,Fan17}, 
but not to phase synchronization.
\end{remark}

Given $S_i$ ($i=1,2$) defined as a parallelogram $(a_i,b_i,e_i)$, 
in order to show the phenomenon of phase synchronization,
we first {\em cover} $S_i$  with a {\em finite} set 
$\{B_{j,i}\}_{j\in J_i}$ of balls $B_{j,i}\subset \mathbb{R}^m$ (\ie{} for $i=1,2$,
$S_i\subset \bigcup_{j\in J_i} B_{j,i}$). %
From 1', 2', (*)  and (**), it follows:

\begin{proposition}\label{proposition:laproposition}
Given a covering $\{B_j\}_{j\in J_i}$ of $S_i$ ($i=1,2$), if, for all $(j_1,j_2)\in J_1\times J_2$,
$PROC1(B_{j_1}\times B_{j_2})$ succeeds, then, for all initial
condition $(x_1^0,x_2^0)  \in S$,  there exists $t \in [kT, (k+1)T)$ such that $(x_1(t),x_2(t)) \in S$. Besides:

$	|phase(x_1(t)) - phase(x_2(t))|  \leq \epsilon + \min(e_1/f_1,e_2/f_2)$,\\
where  $e_i$ is the width of $S_i$, and $f_i=|ord(b_i)-ord(a_i)|$ its height ($i=1,2$).
\end{proposition}

When $\epsilon \ll \min(e_1/f_1,e_2/f_2)$, the final difference of phase between $x_1(t)$ and $x_2(t)$ is practically upper bounded by $\min(e_1/f_1, e_2/f_2)$.
Since, by (H),  $e_i$ is ``small'' w.r.t.\ $f_i$, we know by \cref{proposition:laproposition} that,
if $PROC1$ succeeds for a set of balls covering $S$, then:

For any initial point $(x_1^0,x_2^0)\in S$, 
there exists $t\in [kT,(k+1)T)$ such that
$x_1(t)$ and $x_2(t)$ are {\em almost in phase}. In particular, even if
$|phase(x_1^0)-phase(x_2^0)|\approx 1$ (when $x_1^0$ is located near $a_1$ and $x_2^0$ near $b_2$, or symmetrically), we have:
$|phase(x_1(t))-phase(x_2(t))|\approx 0$.

\section{Example: Brusselator Reaction-Diffusion}\label{section:brusselator}

\begin{figure}[t]
  \centering
  \includegraphics[scale=0.3]{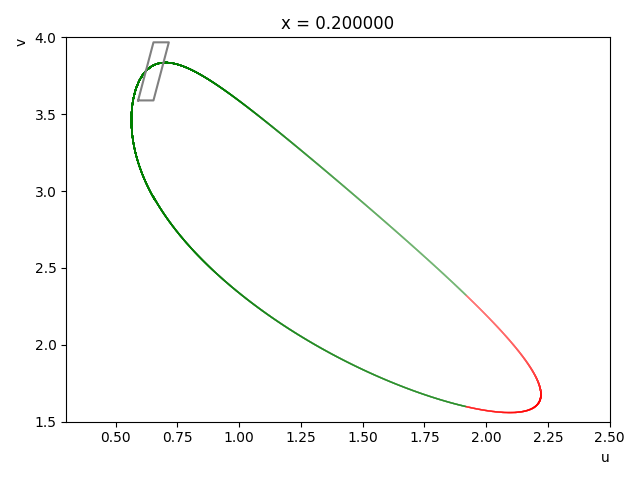}
  \includegraphics[scale=0.3]{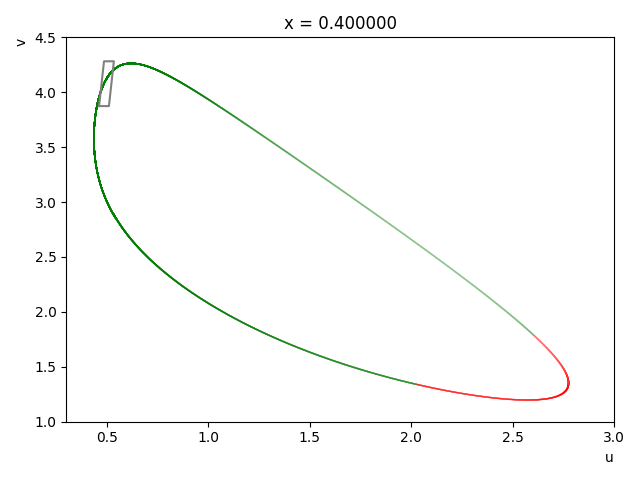}
  \caption{Brusselator: A cyclic trajectory for plan $x=0.2$ (left) and $x=0.4$ (right); the green zone indicates the contractive area ($\lambda < 0$) and the red zone the expansive one ($\lambda > 0$)}
\label{fig:Pouchol2lambda}
\end{figure}

\begin{figure}[t]
  \centering
    \includegraphics[scale=0.44]{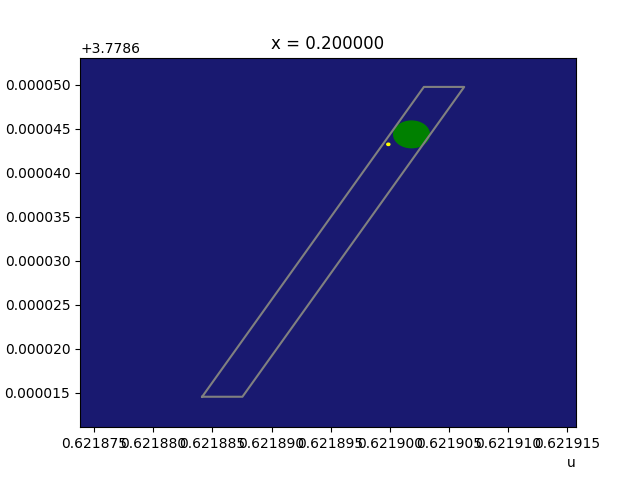}
	\includegraphics[scale=0.44]{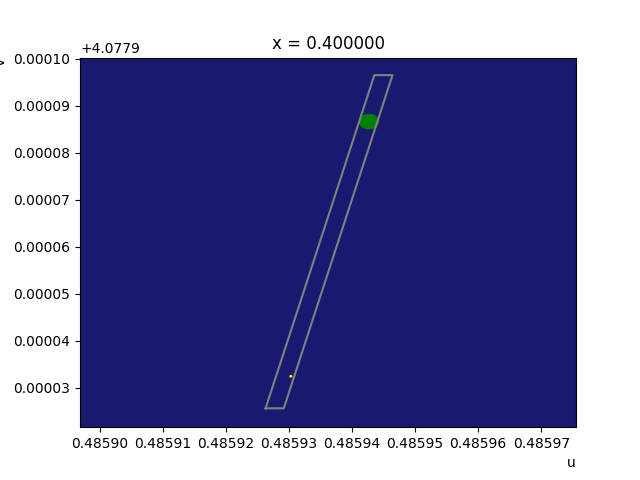}
  \caption{Brusselator: Synchronization of  the two components of a ball, located initially near opposite vertices of the parallelograms (yellow), after $k=5$ periods (green).}
  \label{fig:Pouchol0bis}
\end{figure}

We consider the 1D Brusselator partial differential equation (PDE),
as given in \cite{Chartier93}.
Here we consider  a state of the form $x(y,t)=(u(y,t),v(y,t))$ 
where $y\in \Omega = [0,\ell]$ is the spatial location. The PDE is of the form
\begin{equation}%
\begin{cases}
\frac{\partial u}{\partial t}  = A+u^2v-(B+1)u+\sigma \nabla^2 u\\
\frac{\partial v}{\partial t} = Bu-u^2v+\sigma \nabla^2 v
\end{cases}
\label{Brusselator}
\end{equation}
with boundary condition: $u(0,t)=u(\ell,t)=1$, $v(0,t)=v(\ell,t)=3$,\\
and initial condition: $x_0(y)=(u(y,0),v(y,0))$ with
$u(y,0)=1+sin(2\pi y)$, $v(y,0)=3$.\\
Let: $A=1, B=3, \sigma=1/40$, $\ell =1$.
We transform the PDE into a system of ODEs
by spatial discretization using
a grid of $N+1$ points with $N=4$
(i.e.: $y_i=\frac{i\ell}{N+1}=0.2 i$ for $i=1,2,3,4$).
We thus consider that we have $4$ oscillators 
of state $x(y_i,t)=(u(y_i,t),v(y_i,t))$ %
with initial conditions 
$x(y_i,0)=(u(y_i,0),v(y_i,0))$ ($i=1,2,3,4$). 
These oscillators are coupled by a Laplacian matrix accounting for
the continuous diffusion process;%
the size of the resulting global ODE is $N\times n=4\times 2=8$.
The system of ordinary differential equations for this example is described by 
\begin{equation}
\begin{cases}
	\overset{.}{u_1} = A+u_1^2v_1-(B+1)u_1+\sigma (u_0 - 2u_1 + u_2)\\
	\overset{.}{v_1} = Bu_1-u_1^2v_1+\sigma (v_0 - 2v_1 + v_2)\\
	\overset{.}{u_2} = A+u_2^2v_2-(B+1)u_2+\sigma (u_1 - 2u_2 + u_3)\\
	\overset{.}{v_2} = Bu_2-u_2^2v_2+\sigma (v_1 - 2v_2 + v_3)\\
	\overset{.}{u_3} = A+u_3^2v_3-(B+1)u_3+\sigma (u_2 - 2u_3 + u_4)\\
	\overset{.}{v_3} = Bu_3-u_3^2v_3+\sigma (v_2 - 2v_3 + v_4)\\
	\overset{.}{u_4} = A+u_4^2v_4-(B+1)u_4+\sigma (u_3 - 2u_4 + u_5)\\
	\overset{.}{v_4} = Bu_4-u_4^2v_4+\sigma (v_3 - 2v_4 + v_5)
\end{cases}
\label{Brusselator-explicit}
\end{equation}
with $u_0=u_5=1$ and $v_0=v_5=3$.
By using symmetry, we can reduce the problem to plans 
$x=0.2$ and $x=0.4$ ($x=0.6$ coincides with $x=0.4$, and
$x=0.8$ with $x=0.2$).
We give in \cref{fig:Pouchol2lambda} a typical cyclic trajectory in plans $x=0.2$ and $x=0.4$, during one period $T$.
The coordinates of the parallelepiped vertices are
for plan $x=0.2$: 

$((0.621884, 3.778615),$
$(0.621888, 3.778615),$
$(0.621906, 3.778650),$
$(0.621903, 3.778650))$,\\
and for plan $x=0.4$:

$((0.485926, 4.077926),$
$(0.485929, 4.077926),$
$(0.485946, 4.077997),$
$(0.485943, 4.077997))$.\\
These parallepipeds are depicted in \cref{fig:Pouchol0bis}
(and also at magnified scale in \cref{fig:Pouchol2lambda}). 
The time-step used in Euler's method is $\tau =2\cdot 10^{-4}$, and
the period of the system is $T=34564\tau$. The expansion factor of the 
ball radius after one period is $E= 2.12$. The number of periods considered
for synchronization is $k=5$ (so the expansion factor after $k$ periods $= 2.12^{5} \approx 43$). The radius of the balls covering $S$ is $= 3.5 \cdot 10^{-8}$.

In \cref{fig:Pouchol0bis}, we have depicted an  initial ball (yellow)
with a center of coordinate $(0.622, 3.779)$ in plan $x=0.2$, 
and $(0.486, 4.078)$ in plan $x=0.4$; its radius is  $3.5\cdot 10^{-8}$.
After $k=5$ periods, the image of the yellow ball is the green ball
of center $(0.62190185, 3.77864437)$ in plan $x=0.2$, 
and $(0.48594267, 4.07798666)$ in plan $x=0.4$; the radius is now $1.5\cdot 10^{-6}$.
The phase of the initial ball center is $0.82$ in plan $x=0.2$, and $0.09$ in plan $x=0.4$, so the difference of phase  $\Delta(phase(centers))$, at $t=0$,
is $0.73$.
The phase of the image ball center is $0.87461$ in plan $x=0.2$, and $0.87463$ in plan $x=0.4$, so the difference of phase  $\Delta(phase(centers))$, after $k=5$ periods,  is 
now $2 \cdot 10^{-5}\approx 0$.

\begin{figure}[t]
  \centering
    \includegraphics[scale=0.44]{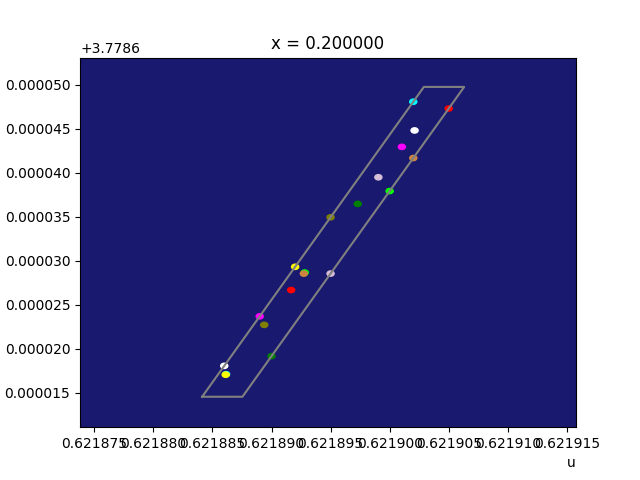}
	\includegraphics[scale=0.44]{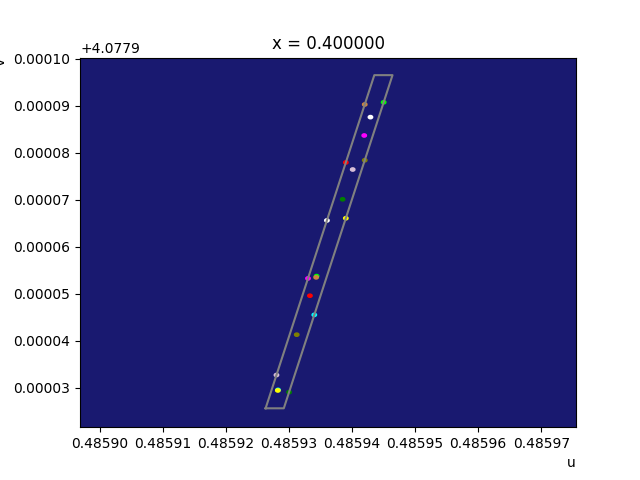}
  \caption{Brusselator: Synchronization of 10 (pairs of) balls,
located initially on the parallelogram perimeters, after $k=5$ periods 
(without radius expansion for clarity).}
  \label{fig:brusselator-simulation-10-points}
\end{figure}

\cref{fig:brusselator-simulation-10-points} depicts 10 (pairs of) initial balls with centers located on the parallelepiped {\em perimeters}, both in plan $x=0.2$ and $x=0.4$.  
The coordinates of the 10 (pairs of) centers, given under the form $(u_1, v_1, u_2, v_2)$, are: 

{\footnotesize
$((0.621890 , 3.778619 ,  0.485930 ,  4.077929 ), $
$( 0.621895 ,  3.778628 ,  0.485928 ,  4.077933 ), $

$( 0.621889 ,  3.778623 ,  0.485933 ,  4.077953 ), $
$( 0.621902 ,  3.778640 ,  0.485934 ,  4.077946 ), $

$( 0.621892 ,  3.778629 ,  0.485939 ,  4.077966 ), $
$( 0.621886 ,  3.778620 ,  0.485936 ,  4.077966 ), $

$( 0.621895 ,  3.778630 ,  0.485942 ,  4.077978 ), $
$( 0.621900 ,  3.778640 ,  0.485945 ,  4.077991 ), $

$( 0.621905 ,  3.778650 ,  0.485939 ,  4.077978 ), $
$( 0.621902 ,  3.778640 ,  0.485942 ,  4.077990 ))$
}

\noindent
After $k=5$ periods, the coordinates of $(u_1,v_1,u_2,v_2)$ become
$(u'_1,v'_1,u'_2,v'_2)$ as follows:

{\footnotesize
$(( 0.621897 , 3.778636 ,   0.485938 ,  4.077970 ),$ 
$(  0.621899 , 3.778639 ,   0.485940 ,  4.077976 ), $

$( 0.621901 ,  3.778643 ,   0.485942 ,  4.077984 ), $
$( 0.621886 ,  3.778617 ,   0.485928 ,  4.077930 ), $

$( 0.621886 ,  3.778617 ,   0.485928 ,  4.077929 ), $
$( 0.621902 ,  3.778645 ,   0.485943 ,  4.077988 ), $

$( 0.621889 ,  3.778623 ,   0.485931 ,  4.077941 ), $
$( 0.621893 ,  3.778629 ,   0.485934 ,  4.077954 ),$ 

$( 0.621892 ,  3.778627 ,   0.485933 ,  4.077950 ), $
$( 0.621893 ,  3.778629 ,   0.485934 ,  4.077953 ))$
}

\noindent
The two components 
$(u_1,v_1)$ and $(u_2,v_2)$ of an initial point, 
as well as the two components $(u'_1,v'_1)$ and $(u'_2,v'_2)$
of its image, are all the 4 represented with the same color
in \cref{fig:brusselator-simulation-10-points}.
The CPU time taken for computing these 10 images is 4,600 seconds 
(for a program\footnote{\label{link}Source codes and figures available at \url{www.lipn.univ-paris13.fr/~jerray/synchro}} of $PROC1$ in Python running on a 2.80 GHz Intel Core i7-4810MQ CPU with 8
GB of memory.).\ea{je donnerais un peu plus de détails sur l'implémentation ; et une URL où la trouver ?}
\jj{url ajoutée pour les deux applications}
\cref{tab:brusselator-phases} gives the phases of the 10 ball centers shown in \cref{fig:brusselator-simulation-10-points}.
After $k=5$ periods, we have  $\Delta(phase(centers))\ll \min(e_1/f_1,e_2/f_2)$, so the difference of phase between the components
of a point starting from {\em anywhere} in a ball (not necessarily from its center) becomes always    $\leq \min(e_1/f_1, e_2/f_2)\approx 0.05$.
The proof has been done here for 10 balls, but should be done for the {\em whole set} of balls covering $S$. It is easy to see that the number of balls covering $S$ is approximatively $\ell_1\ell_2E^{4k}/e_1 e_2$,
where $\ell_i$ is the length of each parallepiped ($i=1,2$).
For example, if $\ell_1/e_1=\ell_2/e_2=20$, $E^k=40$, roughly as in
Brusselator, the number of balls is $400\times 40^4=2^{10}\cdot 10^6\approx 10^9$, which is huge.
However the analysis can be {\em decomposed} into $k$
periods, and accessibility per period proven {\em separately} from one intermediate area to the next, thus exponentially decreasing the number of balls.
In this case, the procedure has to be performed successively $k$ times, but the number of balls  at each time is now just
$\ell_1\ell_2E^{4}/e_1 e_2$, which is  $400\times 2^4=6400$.

\begin{table}[ht]
\caption{The list of phases of 10 ball centers for the Brusselator example.}
\centering

{\footnotesize
\begin{tabular}{lrrrrrr} 
\toprule
Phases\\  
\midrule 
Point &phase initial&phase initial&phase image & phase image & $\Delta(phase(centers))$ & $\Delta(phase(centers))$\\ 
 & point in $u_1$ & point in $u_2$ & point in $u_1$  & point in $u_2$ & for initial point & for image point \\
\midrule 
1 & $0.13$  &  $0.05$  &  $0.63224$  &  $0.63221$  &  $0.08$  &  $2\cdot 10^-5$\\
2 & $0.40$  &  $0.10$  &  $0.72512$  &  $0.72511$  &  $0.30$  &  $8\cdot 10^-6$\\
3 & $0.26$  &  $0.39$  &  $0.83112$  &  $0.83113$  &  $0.13$  &  $6\cdot 10^-6$\\ 
4 & $0.95$  &  $0.28$  &  $0.0383$  &  $0.0382$  &  $0.67$  &  $9\cdot 10^-5$\\
5 & $0.42$  &  $0.57$  &  $0.0366$  &  $0.0365$  &  $0.15$  &  $9\cdot 10^-5$\\
6 & $0.10$  &  $0.56$  &  $0.88834$  &  $0.88836$  &  $0.46$  &  $1\cdot 10^-5$\\
7 & $0.58$  &  $0.74$  &  $0.2103$  &  $0.2102$  &  $0.16$  &  $7\cdot 10^-5$ \\
8 & $0.66$  &  $0.92$  &  $0.3929$  &  $0.3928$  &  $0.25$  &  $5\cdot 10^-5$ \\
9 & $0.93$  &  $0.74$  &  $0.3318$  &  $0.3317$  &  $0.19$  &  $6\cdot 10^-5$ \\
10 & $0.77$  &  $0.91$  &  $0.3890$  &  $0.3889$  &  $0.14$  &  $5\cdot 10^-5$ \\
\bottomrule
\end{tabular}
}
\label{tab:brusselator-phases}
\end{table}

\section{Example: Passive biped model}\label{section:biped}
\begin{figure}[t]
  \centering
  \includegraphics[scale=0.3]{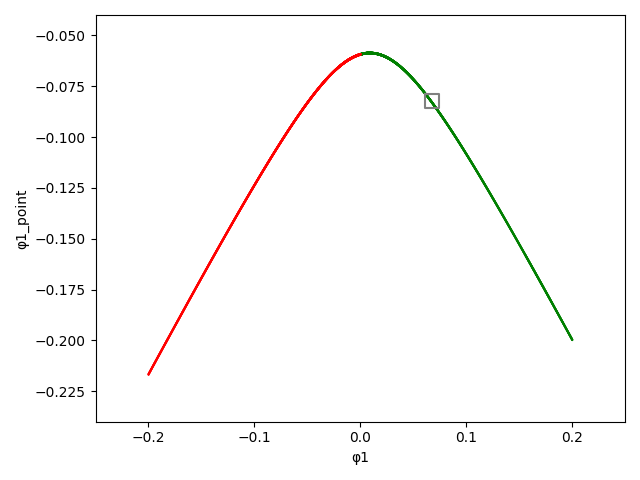}
  \includegraphics[scale=0.3]{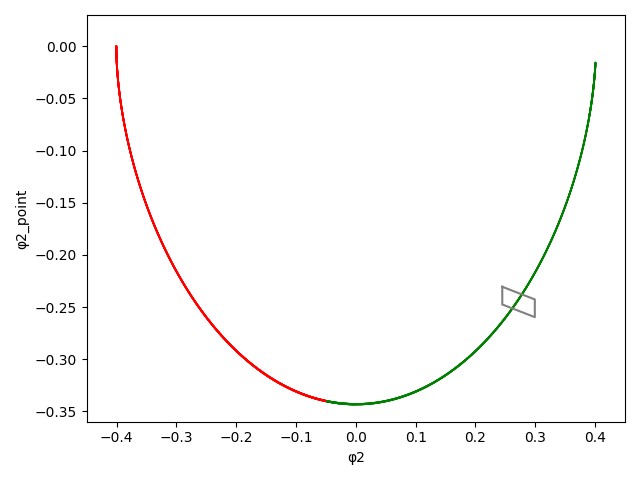}
  \caption{Biped: A cyclic trajectory for plan $\phi_1$ (left) and $\phi_2$ (right); the green zone indicates the contractive area ($\lambda < 0$) and the red zone the expansive one ($\lambda > 0$)}
\label{fig:biped-lambda}
\end{figure}

So far, we he have considered only {\em continuous systems} governed by ODEs.
It is possible to extend the method of verification of phase synchronization to
{\em hybrid systems}, \ie{} continuous systems which, upon the satisfaction
of a certain {\em state condition} (``guard''), may {\em reset} instantaneously
the state before resuming the application of ODEs.
Many works in the domain of symbolic control have explained how to
compute an overapproximation of the {\em intersection} of
the current set of reachability with the guard condition, and perform
the reset operation (see, \eg{} \cite{GirardG08,AlthoffK12,KochdumperA20}).
Our symbolic Euler's  method can be extended along these lines without major 
problems. We describe here the results of such an extension to the {\em passive biped model} \cite{McGeer90}, seen as a hybrid oscillator.
The passive biped model exhibits indeed a stable limit-cycle oscillation for appropriate parameter values that corresponds to periodic movements of the legs \cite{Nakao17}.
The model has a continuous state variable $\textbf{\textit{x}}(t) = (\phi_1(t), \overset{.}{\phi_1}(t), \phi_2 (t), \overset{.}{\phi_2} (t))^\top$. The dynamics is described by $\dot{\textbf{\textit{x}}}=\textbf{{\em f}}(\textbf{\textit{x}})$ with:
\begin{equation}
\textit{\textbf{f}}(\textbf{\textit{x}}) = \begin{pmatrix}
    \overset{.}{\phi_1} \\
    \sin(\phi_1-\gamma)\\
    \overset{.}{\phi_2}\\
    \sin(\phi_1 - \gamma) + \overset{.}{\phi_1^{2}} \sin \phi_2 - \cos (\phi_1 - \gamma) \sin \phi_2
  \end{pmatrix}
\label{Biped_F}
\end{equation}
\begin{equation}
Reset (\textbf{\textit{x}})  = \begin{pmatrix}
    -\phi_1\\  
    \overset{.}{\phi_1}\sin(2\phi_1)\\
    -2\phi_1\\
    \overset{.}{\phi_1 }\cos 2\phi_1 ( 1 - \cos 2\phi_1 )
  \end{pmatrix}
\label{Biped_phi}
\end{equation}
\begin{equation}
Guard (\textbf{\textit{x}}) \equiv (2\phi_1 - \phi_2 =0\ \wedge \phi_2<-\delta) .
\label{Biped_pi}
\end{equation}
We set $\delta = 0.1$ and $\gamma = 0.009$. 
See \cite{McGeer90} for details.
We give in \cref{fig:biped-lambda} a typical cyclic trajectory in plans $\phi_1$ and $\phi_2$, during one period~$T$.
The coordinates of the parallelepiped vertices are for plan $\phi_1$:

{\footnotesize
$((0.067939, -0.083172),$
$(0.067943, -0.083172),$
$(0.067943, -0.083169),$
$(0.067939, -0.083169))$,
}

\noindent
and for plan $\phi_2$:

{\footnotesize
$((0.271972, -0.2427 25),$
$(0.271983, -0.2427 34),$
$(0.271983, -0.242731),$
$(0.271972, -0.242722))$.
}

\begin{figure}[t]
  \centering
    \includegraphics[scale=0.31]{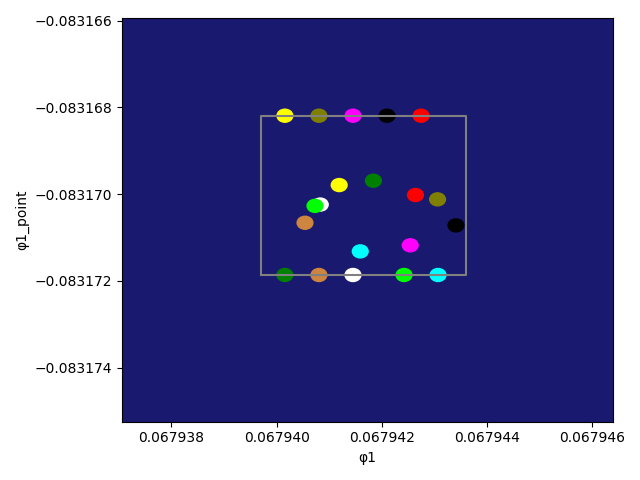}
	\includegraphics[scale=0.31]{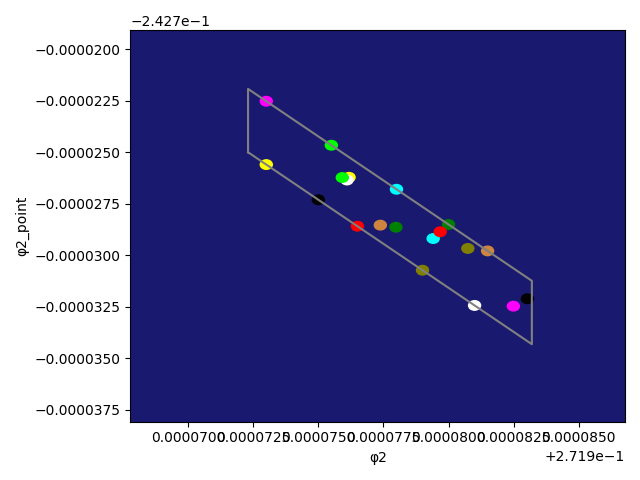}
  \caption{Biped: Synchronization of 10 (pairs of) balls,
located initially on the parallelogram perimeters, after $k=30$ periods 
(without radius expansion for clarity).}
  \label{fig:biped-simulation}
\end{figure}

\noindent
These parallepipeds are depicted in \cref{fig:biped-simulation}
(and also at magnified scale in \cref{fig:biped-lambda}). 
The time-step used in Euler's method is $\tau=2\cdot 10^{-5}$.
The period of the system is $T=776440\tau$.
The radius expansion factor after one period is $E= 2.63$.
The number of periods considered for synchronization is
$k=30$.

\cref{fig:biped-simulation} depicts 10 (pairs of) initial balls with
centers located on the parallelepiped {\em perimeters}, both in plan $\phi_1$ and $\phi_2$.  
The coordinates of these 10 (pairs of) centers, given under the form $(\phi_1, \overset{.}{\phi_1}, \phi_2, \overset{.}{\phi_2})$, are:

{\footnotesize
$((0.067940, -0.083172, 0.27198,   -0.242729), $
$(0.067942,  -0.083168, 0.271975,  -0.242727), $

$(0.067941,  -0.083168, 0.271973,  -0.242723), $
$(0.067943, -0.0831719, 0.271978,  -0.242727), $

$(0.067940,  -0.0831682, 0.271973,  -0.242726), $
$(0.067941,  -0.0831719, 0.271981,  -0.242732), $

$(0.067940,  -0.0831682, 0.271979,  -0.242731), $
$(0.067942, -0.0831719, 0.271976, -0.242725), $

$(0.067943,  -0.0831682, 0.271977, -0.242729), $
$(0.067941,  -0.0831719, 0.271981, -0.242730))$
}

\noindent
The coordinates $(\phi'_1, \overset{.}{\phi'_1}, \phi'_2, \overset{.}{\phi'_2})$
of their images after 30 periods are:

{\footnotesize
$(( 0.06794 18, -0.0831697,  0.2719 78, -0.2427 29), $
$( 0.06794 34, -0.08317 07,  0.2719 83, -0.2427 32), $

$( 0.06794 25, -0.08317 12,  0.2719 82, -0.2427 32), $
$( 0.06794 16, -0.08317 13,  0.2719 79, -0.2427 29), $

$( 0.06794 12, -0.08316 98,  0.2719 76, -0.2427 26), $
$( 0.06794 08, -0.08317 02,  0.2719 76 , -0.2427 26), $

$( 0.06794 31, -0.08317 01,  0.2719 81, -0.2427 30), $
$( 0.06794 07, -0.08317 03,  0.2719 76, -0.2427 26), $

$( 0.06794 26, -0.08317 00,  0.2719 80, -0.2427 29), $
$( 0.06794 05, -0.08317 07,  0.2719 77, -0.2427 29)) $
}

\noindent
The two components 
$(\phi_1,\dot{\phi}_1)$ and $(\phi_2,\dot{\phi}_2)$ of an initial point, 
as well as the two components $(\phi'_1,\dot{\phi}'_1)$ and $(\phi'_2,\dot{\phi}'_2)$
of its image, are all the 4 represented with the same color
in \cref{fig:biped-simulation}.
The CPU time taken for computing the 10 images is 6,800 seconds\ea{redire que c'est avec le même programme ?} (for a program\footnoteref{link} written in Python running on the same machine used for the Brusselator example).
\cref{tab:biped-phases} gives the phases of the 10 (pairs of) points shown in \cref{fig:biped-simulation}. After $k=30$ periods, we have $\Delta(phase(centers))\leq 0.25$.
Since $\min(e_1/f_1,e_2/f_2)\approx 0.15$, 
the difference of phase between the components
of a point starting {\em anywhere} from a ball (not necessarily fom its center), becomes always
$\leq 0.4$. Here again, the proof has been done for 10 balls, but should be done for the whole set of balls covering $S$.

\begin{table}[ht]
\caption{The list of phases of 10 ball centers in the biped example.}
\centering
{\footnotesize
\begin{tabular}{lrrrrrr} 
\toprule
Phases\\  
\midrule 
Point & Phase initial  & Phase initial  & Phase image & phase image  & $\Delta(phase(centers))$ & $\Delta(phase(centers))$ \\ 
 & point in $\phi_1$ & point in $\phi_2$ & point in $\phi_1$  & point in $\phi_2$ & for initial point &  for image point\\
\midrule 
1 & 0.88  &  0.29  &  0.45  &  0.48  &  0.59  &  0.03 \\
2 & 0.38  &  0.75  &  0.05  &  0.02  &  0.37  &  0.03 \\
3 & 0.55  &  0.94  &  0.27  &  0.07  &  0.39  &  0.21 \\
4 & 0.14  &  0.48  &  0.52  &  0.35  &  0.34  &  0.17 \\
5 & 0.88  &  0.94  &  0.62  &  0.64  &  0.05  &  0.03 \\
6 & 0.55  &  0.20  &  0.71  &  0.65  &  0.35  &  0.06 \\
7 & 0.72 &  0.39  &  0.14  &  0.23  &  0.33  &  0.09 \\
8 & 0.30  &  0.71  &  0.74  &  0.67  &  0.40  &  0.07 \\
9 & 0.22  &  0.61  &  0.25  &  0.32  &  0.40  &  0.08 \\
10 & 0.72 &  0.16  &  0.78  &  0.53  &  0.56  &  0.25 \\
\bottomrule
\end{tabular}
}
\label{tab:biped-phases}
\end{table}

\section{Final Remarks}\label{section:conclusion}
We have described a symbolic reachability method to prove phase 
synchronization of oscillators, and illustrated it on the
Brusselator and biped examples.
The method is inspired by the classical ``direct method'' which shows that
a {\em finite} number of points, displaced from their original position on a synchronization orbit, return after some time into a close neighborhood of
the orbit. In contrast to the classical method, our symbolic method
shows an analogous property for the {\em infinite} set $S$ of points located around a portion of the orbit. Such a set $S$ can be determined using simulation methods, but we assume here that it is given.
Note that our method  guarantees that the solution components
are {\em almost} synchronized when they pass into~$S$, whereas standard synchronization states the stronger property of {\em convergence} to the synchronization orbit.

Because of the magnification of the balls on a non-contractive space ($\lambda>0$), one is forced to start with small initial balls, and the coverage of $S$ requires a priori a huge number of balls. However, as explained on the Brusselator example, the analysis can be decomposed into 
periods, and accessibility per period proven {\em separately} from one intermediate area to the next, thus exponentially decreasing the number of balls. Note that the ball magnification problem does {\em not} occur on a {\em contractive} system ($\lambda <0$), \eg{} for Brusselator with a {\em large diffusion coefficient} $\sigma$, so the reachability analysis is easier in this case.

We focused here on $n=2$ components
with state space dimension $m=2$. The extension to
$n, m \geq 3$
is easy in principle, but causes combinatorial explosion of the number of 
balls covering~$S$.
In order to solve this ``curse of dimensionality'', it would be interesting in future work to adapt the
classical ``adjoint'' method (or {\em phase reduction} \cite{Nakao17})  rather
than the ``direct'' method used here.
Note also that our guaranteed method of phase synchronization 
can be used with any symbolic reachability procedure other than Euler's method.

\newcommand{\LNCS}{LNCS}

\ifdefined\VersionAuthor
	\renewcommand*{\bibfont}{\small}
	\printbibliography[title={References}]
\else
	\bibliographystyle{plain} %
	\bibliography{arch}
\fi

\end{document}